# Superconductivity in undoped $T'$ cuprates with $T_c$ over 30 K

Osamu Matsumoto [a], Aya Utsuki [a], Akio Tsukada [b], Hideki Yamamoto [c], Takaaki Manabe [d], and *Michio Naito [a]

[a] Department of Applied Physics, Tokyo University of Agriculture and Technology

Naka-cho 2-24-16, Koganei, Tokyo 184-8588, Japan

[b] Geballe Laboratory for Advanced Materials, Stanford University,

Stanford, California 94305, USA

[c] NTT Basic Research Laboratories, NTT Corporation, 3-1 Morinosato-Wakamiya, Atsugi,

Kanagawa 243-0198, Japan

[d] National Institute of Advanced Industrial Science and Technology (AIST)

Higashi 1-1-1, Tsukuba, Ibaraki 305-8565, Japan

**Abstract**

Undoped cuprates have long been considered to be antiferromagnetic insulators. In this article, however, we report that superconductivity is achieved in undoped $T'$-$RE_2CuO_4$ ($RE$ = Pr, Nd, Sm, Eu, and Gd). Our discovery was performed by using metal-organic decomposition (MOD), an inexpensive and easy-to-implement thin-film process. The keys to prepare the superconducting films are firing with low partial-pressure of oxygen and reduction at low temperatures. The highest $T_c$ of undoped $T'$-$RE_2CuO_4$ is over 30 K, substantially higher than "electron-doped" analogs. Remarkably, $Gd_2CuO_4$, even the derivatives of which have not shown superconductivity so far, gets superconducting with $T_c^{onset}$ as high as ~ 20 K. The implication of our discovery is briefly discussed.

*) Corresponding author. Address: Department of Applied Physics, Tokyo University of Agriculture and Technology, Naka-cho 2-24-16, Koganei, Tokyo 184-8588, Japan. Tel. +81 42 388 7229; fax: +81 42 385 6255. E-mail address: minaito@cc.tuat.ac.jp.

**Introduction**

Undoped cuprates have long been considered to be antiferromagnetic Mott insulators (charge-transfer insulators). For example, $La_2CuO_4$ with the $K_2NiF_4$ (abbreviated as $T$) structure is an insulator with no doubt. $RE_2CuO_4$ ($RE$: rare-earth element) with the $Nd_2CuO_4$ (abbreviated as $T'$) structure has also been believed as a Mott insulator since the discovery of "electron-doped" superconductors, $T'$-$(RE,Ce)_2CuO_4$ in 1989 [1]. In this article, however, we report undoped $T'$-$RE_2CuO_4$ to be superconducting. The origin of the sharp contradiction between the past and our results can be traced to impurity oxygen ($O_{ap}$) at the apical site. Empirically, apical oxygen atoms in $T'$ cuprates play the role of very strong scatterer as well as pair-breaker, presumably due to magnetic (Kondo) scattering by $O_{ap}$-induced Cu spins [2]. Therefore the generic phase diagram of the $T'$-cuprates can be reached only after removing apical oxygen atoms completely. The first report along this line was made by Brinkmann *et al*. in 1995 [3]. They reported that the superconducting (SC) window for $Pr_{2-x}Ce_xCuO_4$ expands by an improved reduction process down to $x = 0.04$ even with slightly increasing $T_c$. The expanded SC window can be explained by the suppression of $O_{ap}$-induced antiferromagnetic (AF) order. Employing a thin film process, we also demonstrated that the superconducting range in $T'$-$La_{2-x}Ce_xCuO_4$ expands down to $x = 0.045$ [4], and also that the end-member $T'$-$La_2CuO_4$, which has been supposed to be a Mott insulator, shows metallic resistivity down to 150 K with $\rho(300\ K)$ as low as 2 mΩcm [5]. Furthermore, in the previous article, we reported the superconductivity with $T_c \sim 25$ K in "undoped" $T'$-$(La,RE)_2CuO_4$, which was achieved by state-of-the-art molecular beam epitaxy (MBE) technique [6]. However, there has been a great controversy on the superconducting nature in $T'$-$(La,RE)_2CuO_4$: truly undoped or electron-doped due to oxygen deficiencies. One drawback to obtaining definitive conclusion is that $T'$-$(La,RE)_2CuO_4$ is difficult to synthesize by bulk processes. Therefore they are not amenable to any measurement of oxygen nonstoichiometry like chemical analysis or neutron diffraction. Hence only speculative discussions have been made based on transport measurements such as Hall coefficient and Nernst signal [7, 8].

In this article, we report the superconductivity with $T_c$ over 30 K in $T'$-$RE_2CuO_4$ ($RE$ = Pr, Nd, Sm, Eu, and Gd). This was achieved by metal-organic deposition (MOD), a rather inexpensive and easy-to-implement thin film process with no fundamental difference from bulk synthesis, which is in contrast to leading-edge MBE required to produce the first generation of undoped superconductors, $T'$-$(La,RE)_2CuO_4$. The key recipes to achieve superconductivity in undoped $T'$-$RE_2CuO_4$ by MOD are low-$P_{O2}$ firing and subsequent low-temperature reduction, which lead, we believe, to cleaning up most of apical oxygen atoms harmful to high-$T_c$ superconductivity. Some preliminary results have been announced in conference articles. [9, 10]

**Experimental**

The superconducting $T'$-$RE_2CuO_4$ thin films were prepared by the MOD method by using $RE$ and Cu naphtenate solutions [11]. The solutions were mixed with the stoichiometric ratio and the mixed solution was spin-coated on substrates by 3000 rpm for 30 sec. As a substrate, we used $SrTiO_3$ (STO) (100) and $DyScO_3$ (DSO) (110) [12]. The coated films were first calcined at 400°C in air to obtain precursors. Then the precursors were fired at 850 – 900°C in a tubular furnace under a mixture of $O_2$ and $N_2$, varying the oxygen partial pressure $P_{O2}$ from 2 x $10^{-4}$ atm to 1 atm. Finally the films were "reduced" in vacuum (<$10^{-4}$ Torr ≈ $10^{-7}$ atm) at various temperatures for $O_{ap}$ removal. Films with no reduction were also prepared for a reference, and named as "as-grown". After the reduction, the films were furnace-cooled to ambient temperatures in vacuum to avoid re-absorption of oxygen atoms, which is a quite important step. The thickness of resultant films was typically 500 – 1000 Å, although it was not easy to determine accurately. It is because $T'$-$RE_2CuO_4$ is fairly acid-resistive and is difficult to make steep edge for profilometry. The crystal structure and the *c*-axis lattice parameter ($c_0$) of the films were determined with a 2-circle X-ray diffractometer. The resistivity was measured by a standard four probe method, and the Hall coefficient by a five-probe method.

**Results**

Figure 1 shows the X-ray diffraction (XRD) patterns of typical films prepared by MOD. All peaks are sharp and can be indexed to (00*l*) of the $Nd_2CuO_4$ structure, indicating that the films are single-phase *T'*, and also single-crystalline as achieved via solid-state epitaxy. The peaks shift systematically toward larger $2\theta$ from Pr to Gd, indicating that the *c*-axis shortens in this order.

The end-member *T'* cuprates are semiconducting with standard bulk synthesis, as well known. In contrast, they become fairly metallic after "reduction" in thin film form, especially for large $RE^{3+}$ ions like La, Pr, Nd and Sm [6], which is probably because thin films are advantageous due to a large surface-to-volume ratio in removing apical oxygen atoms. However, the optimization of post-reduction conditions alone is not sufficient to obtain superconductivity in the end-member compounds. The key step for superconductivity is to fire films in a low-$P_{O2}$ atmosphere, the aim of which is to minimize the amount of impurity oxygen in advance before the post-reduction process. Figure 2 shows the effect of $P_{O2}$ during firing, which compares the resistivity of two $Sm_2CuO_4$ films (A and B): film A fired at 900°C for 1 hour in $P_{O2}$ = 1 atm, followed by reduction at 750°C for 10 min, and film B fired at 850°C for 1 hour in $P_{O2}$ = 2.8 x $10^{-3}$ atm, followed by reduction at 440°C for 10 min. Film A is metallic down to 180 K with $\rho$(300 K) ~ 100 mΩcm, but shows upturn at lower temperatures. By contrast, film B is all the way metallic with $\rho$(300 K) as low as 900 μΩcm, and shows superconductivity at $T_c^{onset}$ = 28 K ($T_c^{end}$ = 25 K). The dramatic effect of low-$P_{O2}$ synthesis is actually corroborated from the structural aspect. The *c*-axis lattice constant ($c_0$) are 11.97 Å (11.975 Å before reduction) for film A and 11.94 Å (11.955 Å before reduction) for film B.  The $c_0$ of film A shows a good agreement with the previously reported bulk value [13] whereas the $c_0$ for film B is substantially (~0.03 Å) shorter.  Taking account of the well-established trend that the $c_0$ increases with the amount of $O_{ap}$ atoms [14], our experimental results indicate that low-$P_{O2}$ synthesis is quite effective in preventing $O_{ap}$ atoms from being introduced to lattice.

Figures 3 show the reduction condition dependence of the film properties for $Pr_2CuO_4$ films.  The data were taken from the films fired with identical conditions (850°C for 1h in $P_{O2}$ = 2.8 x 10$^{-4}$ atm) but reduced differently, namely either with a different reduction temperature ($T_{red}$) for Fig. 3(a) or a different duration ($t_{red}$) for Fig. 3(b).  The "as-grown" film (not shown) is semiconducting (d$\rho$/dT < 0) and shows no superconductivity. As shown in Fig. 3(a), the reduction with $T_{red}$ = 420°C and $t_{red}$ = 10 min is sufficient for the film to be superconducting ($T_c^{onset}$ ~ 30 K).  Increasing $T_{red}$ up to 450°C lowers the resistivity and sharpens the superconducting transition.  Further increase of $T_{red}$, however, deteriorates the film properties.  Similar behavior is observed in Fig. 3(b) by varying $t_{red}$ with $T_{red}$ fixed at 450°C.  A long transition tail appears already with $t_{red}$ = 15 min and superconductivity disappears with $t_{red}$ = 60 min although there is almost no trace of decomposition as judged from the XRD pattern.  It is most likely that excessive reduction leads to loss of oxygen atoms in the $CuO_2$ planes, as supported from the oxygen nonstoichiometry experiments [15].  The reduction with $T_{red}$ = 450°C and $t_{red}$ = 5 – 10 min provides $Pr_2CuO_4$ films with optimal properties, namely highest $T_c$ ($T_c^{onset}$ = 31.5 K, $T_c^{end}$ = 27.3 K) and lowest resistivity [$\rho$(300 K) ~ 400 $\mu\Omega$cm and $\rho$ ($T_c^+$) ~ 50 $\mu\Omega$cm].  The low resistivity value, which is only about twice as high as our best MBE-grown $Pr_{1.85}Ce_{0.15}CuO_4$ films [16], guarantees the high-quality of MOD films.

It is difficult to make a quantitative evaluation of the amount of $O_{ap}$ atoms remaining in thin films, but we have found that one good measure for it may be the c-axis lattice constant ($c_0$).  Figures 4 plots the room-temperature resistivity $\rho$(300 K) (lower) and $T_c$ (upper) as a function of $c_0$ for all of our MOD-grown $Pr_2CuO_4$ films. The $\rho$(300 K) decreases rapidly as $c_0$ decreases from the bulk value (broken line), and has a minimum around $c_0 \sim 12.18$ Å. A further decrease in $c_0$ leads to a gradual increase of $\rho$(300 K). We can divide a range of $c_0$ into three regions:

(I) *Insufficient reduction* (12.23 Å ≥ $c_0$ ≥ 12.20 Å)

The $\rho$(300 K) decreases with decrease of $c_0$. Superconductivity with $T_c^{onset} \sim 25$ K suddenly appears for $c_0 < 12.22$ Å. Decreasing $c_0$ towards 12.20 Å, the $T_c^{onset}$ gradually increases to ~ 30 K, and the $T_c^{end}$ also improves. The temperature dependence of resistivity for $Pr_2CuO_4$ has upturn at low temperature, which is a feature specific to *insufficient reduction*.

(II) *Optimum reduction* (12.20 Å ≥ $c_0$ ≥ 12.19 Å)

The $\rho$(300 K) still decreases with $c_0$, and the optimum superconductivity is obtained with $T_c^{onset} > 30$ K and $T_c^{end} > 27$ K.

(III) *Excessive reduction* (12.19 Å ≥ $c_0$ ≥ 12.16 Å)

The $\rho$(300 K) is still decreasing with $c_0$ until 12.18 Å, but the $T_c^{onset}$ starts to decrease and the superconducting transition becomes broad with $T_c^{end}$ below 4.2 K. With a further decrease of $c_0$ below 12.18 Å, the resistivity gradually increases and superconductivity eventually disappears. One feature contrast to that for region (I) is that the temperature dependence of resistivity is metallic in the whole temperature range without low-temperature upturn.

Figures 3 and 4 seem to indicate that $O_{ap}$ removal proceeds together with loss of oxygen [O(1)] in the $CuO_2$ planes. The predominant effect in region (I) is removal of $O_{ap}$, which improves the film properties. In contrast, the predominant in region (III) is loss of O(1), which degrades the film properties although $O_{ap}$ may still being removed as judged from the shrinkage of $c_0$. The removal of $O_{ap}$ proceeds slightly quicker than the loss of O(1), which provides the stage potential for high-$T_c$ superconductivity with $O_{ap}$ atoms mostly removed and still with O(1) atoms mostly preserved as in region (II).

Figure 5 shows the reduction dependence of the Hall coefficient ($R_H$). The films are labeled from A to F according to the reduction strength, as correspondingly indicated in Fig. 4. The Hall coefficient ($R_H$) changes the sign with reduction: negative for insufficiently reduced films and positive for excessively reduced films. The optimally reduced film (film C) has nearly $R_H \sim 0$. The data in Fig. 5 qualitatively resembles the reduction dependence of $\rho$-$T$ and $R_H$ in "optimally doped" $T'$-Nd$_{1.85}$Ce$_{0.15}$CuO$_4$ reported by Jiang *et al*. [17], especially in that $R_H$ goes up with the change of sign from negative to positive with reduction. In general, the Hall coefficients in metals are a very complicated function of anisotropic Fermi velocity as well as anisotropic relaxation time, which makes the interpretation of the data non-straightforward [18]. The simplistic two-band model, although neither quantitative nor realistic, gives some insight for the sign change of $R_H$.

$$R_H = \frac{\mu_p \sigma_p - \mu_n \sigma_n}{(\sigma_p + \sigma_n)^2} = \frac{pe\mu_p^2 - ne\mu_n^2}{(pe\mu_p + ne\mu_n)^2}$$

The numerator that determines the sign of $R_H$ is a mobility-weighed sum (sign-counted) of the conductivities of the two bands. Then, even the minority carrier can determines the sign of $R_H$ if the minority carriers are much more mobile than the majority carriers. Based on this model, the behavior of $R_H$ in Fig. 5 can be interpreted as follows. We assume that the majority carriers are hole and the minority carriers are electron, and also that O$_{ap}$ scatters mainly hole carriers, which is consistent with the recent analytical articles on the transport properties of Pr$_{2-x}$Ce$_x$CuO$_4$ films [19, 20]. With a fair amount of O$_{ap}$ atoms (insufficiently reduced films), the hole mobility $\mu_p$ is low and $R_H$ is determined by the minority carriers, namely negative. Cleaning up O$_{ap}$ atoms, the hole mobility $\mu_p$ improves then $R_H$ becomes positive and approaches to $\frac{1}{pe}$ [21]. The observed $R_H$ of film D to F is 5 to 10 x 10$^{-10}$ m$^3$/C, which gives $p$ of 1.25 to 0.625 x 10$^{22}$ holes/cm$^3$, which amounts to 1.2 – 0.6 holes/Cu, namely almost half-filled.

Table I is a summary of the *RE*-dependence. All *RE*'s we have tested show superconductivity except for *RE* = La, which does not form the *T'* structure by MOD [24]. It should be noted that undoped $T_c$ is substantially higher than electron-doped $T_c$. The highest $T_c^{onset}$ is 32.5 K in $Nd_2CuO_4$. The most remarkable is $Gd_2CuO_4$: there has been no report [25] that electron-doped $Gd_{2-x}Ce_xCuO_4$ gets superconducting, whereas undoped $Gd_2CuO_4$ in the present study has $T_c^{onset}$ as high as 19.0 K. The $T_c$ of electron-doped *T'* shows a steep dependence on *RE* [23], in contrast, the $T_c^{onset}$ of undoped *T'* is rather flat around 30 K except for Gd. Lower $T_c$ in *RE* = Gd is not intrinsic and most likely due to the material problem that $O_{ap}$ removal is difficult for smaller $RE^{3+}$. We are aware that currently obtained $T_c$ of undoped *T'* is substantially process-dependent.

**Discussion**

The key question is the nature of superconductivity in the *nominally undoped T'-RE$_2$CuO$_4$*. Are these materials (a) hole-doped or (b) electron-doped or (c) self-doped or (d) truly undoped? At first, it is safe to exclude hole-doped superconductivity due to excess oxygen. It is well-known that excess oxygen induces hole-doped superconductivity in *T*-$La_2CuO_4$ [26], but not in *T'*-$RE_2CuO_4$ [27]. With regard to electron-doped superconductivity, more careful examinations are required. There are two regular oxygen sites in the *T'* structure: O(1) in the $CuO_2$ planes and O(2) in the fluorite $RE_2O_2$ planes. Unfortunately, even with current state-of-the-art technologies it is impossible to determine directly the oxygen content of a thin film within the accuracy of a few percent, and hence one has to rely somewhat on speculations. The first possibility is the oxygen [O(1)] deficiencies in the $CuO_2$ layers. The O(1) deficiencies actually exist especially in excessively reduced films and probably even in films with the highest $T_c$, but imperfect $CuO_2$ planes are unfavorable to achieving high-$T_c$ superconductivity. The second possibility is the oxygen [O(2)] deficiencies in the fluorite $RE_2O_2$ layers, which one can imagine most easily when one sees the resemblance of the reduction dependence in the present article with the well-known Ce-doping dependence. (ref: The resemblance can be traced to the Ce-doping dependent complex oxygen chemistry in *T'* cuprates). However, the oxygen [O(2)] loss in the fluorite $RE_2O_2$ planes is unlikely to occur since solid-state chemistry tells us that the *RE*-O bond is much stronger than the Cu-O bond [28], which can be realized by the fact that the phase stability line ($P_{O2}$-vs-1/$T$) of *T'* cuprates coincides almost with that of CuO, but is quite far above that of $RE_2O_3$ [10]. In fact, there is no neutron diffraction experiment [29, 30] showing any change in the O(2) occupancy with reduction even though some of early experiments showed that the O(2) occupancy is not full (namely, slightly below 2.00). This indicates that the reduction dependence of physical properties in *T'*-$RE_2CuO_4$ are not explained by the change in the O(2) occupancy. Furthermore the observation of more positive Hall coefficient with reduction also argues against a simple electron doping mechanism via the O(2) deficiency.

On the third possibility of self-doping, there is a classic example of $Tl_2Ba_2Ca_{n-1}Cu_nO_{2n+4}$ family [31]. They are nominally undoped if one assumes the valence of Tl as 3+, but in reality they are hole-doped. This is because the Tl 6*s* band in $Tl_2Ba_2Ca_{n-1}Cu_nO_{2n+4}$ is significantly below the Fermi level so that it has a large electron pocket removing electrons from the Cu-O *dp*σ antibonding band, which yields self-doping ($Tl^{3+} + Cu^{2+} \leftrightarrow Tl^{(3-\delta)+} + Cu^{2+\delta}$). Another example is recently discovered four-layered $Ba_2Ca_3Cu_4O_8F_2$ with $T_c \sim 60$ K [32]. In this compound, it has been proposed that electrons are transferred from the outer (*o*) pyramidal $CuO_2$ layers to the inner (*i*) square-planar $CuO_2$ layers due to the Madelung potential difference of 0.4 eV, thereby inducing hole-doped superconductivity in the *o* layers and electron-doped superconductivity in the *i* layers. In the case of *T'*-$RE_2CuO_4$, the early band calculation by Massida *et al*. [33] pointed out a few features specific to the *T'* structure, in spite of most of the features common to other HTSC's.

The O(2)$p_x$-$p_y$ bands have the same symmetry as the Cu$d_{x2-y2}$-O(1)$p_x$-$p_y$ and there are vertical O(1)-O(2)-O(1)-O(2) chains.  Both of them push up the anti-bonding O(1)$p_z$-O(2)$p_z$ state to close to $E_F$, leading to that the O(2)$p$ bands lie just below $E_F$.  If the O(2)$p$ bands cross $E_F$, it leads to self-doping in the form of $(RE_2O_2)^{2+\delta}(CuO_2)^{-2-\delta}$ and makes the $RE_2O_2$ charge reservoir block metallic.  However, there has so far been no report that $RE_2O_2$ layers are metallic, and hence self-doping may also be unlikely.

With regard to the final possibility, our Madelung potential calculations [34] demonstrate that the unscreened value ($\Delta_0$) of the charge transfer energy for $T'$-$RE_2CuO_4$ with large $RE^{3+}$ ions to excite an electron from O$^{2-}$ to a neighboring Cu$^{2+}$ is 10-11 eV, significantly (~3 – 4 eV) lower than $\Delta_0$ = 13.7 eV for $T$-La$_2$CuO$_4$, due to the absence of apical oxygen and also larger Cu-O bond length.  The empirical critical value ($\Delta_0^{cr}$) to separate metallic and insulating oxides is 10 eV (Torrance's criterion [35]).  This indicates that the charge transfer gap may close in $T'$-$RE_2CuO_4$.  For such low-$\Delta$ metal, the ionic picture loses its meaning, and the band picture may be favored.  The physics behind the new observation of superconductivity in the nominally undoped $T'$-superconductors at present is not yet fully understood.  However, the amount of O$_{ap}$ atoms seems to affect the phase diagram severely and the occurrence of the "Mott"-like insulating state.  At present, we are thinking that the most likely is undoped superconductivity, which should be established in future stringent experiments.  Since MOD is not fundamentally different from bulk synthesis, superconducting $T'$-$RE_2CuO_4$ may be obtainable in bulk form, which enables oxygen nonstoichiometry measurements such as chemical analysis, neutron diffraction and so on. The electronic state can also be probed by X-ray absorption spectroscopy, photoemission spectroscopy, *etc*.

**Summary**

We discovered superconductivity in $T'$-$RE_2CuO_4$ ($RE$ = Pr, Nd, Sm, Eu, and Gd),

which have been believed as a Mott insulator. The synthesis is rather simple and inexpensive, namely low-$P_{O2}$ firing and subsequent low-temperature reduction of the MOD films. One point to be emphasized is that low-$P_{O2}$ phase field has been almost unexplored in the search for new cuprate superconductors because of the belief that high $P_{O2}$ *should* be required in the synthesis of $Cu^{2+}$ compounds. The highest $T_c$ of undoped $T'$-$RE_2CuO_4$ is over 30 K, substantially higher than "electron-doped" analogs. It is the most likely that these superconductors are truly undoped although this has to be established in future works.

**Acknowledgements**

The authors thank Dr. Y. Krockenberger and Dr. J. Shimoyama for stimulating discussions, and Dr. T. Kumagai for support and encouragement. They also thank Crystec GmbH, Germany for developing new $RE$ScO$_3$ substrates. HY thanks Dr. M. Kasu, Dr. K. Torimitsu, Dr. T. Makimoto, and Dr. J. Yumoto for their support and encouragement. The work was supported by KAKENHI B (18340098) from Japan Society for the Promotion of Science (JSPS).

Alternatively, it seems that a comparison of bond energy between CuO (~270 kJ/mol) and *RE*-O (~420 – 800 kJ/mol) molecules [J. A. Kerr, in: *CRC Handbook of Chemistry and Physics*, 66th ed., CRC Press Inc., Florida, 1985, p.F174] may give more accurate clue which oxygen site is easier to be deficit.

**Figure captions**

Figure 1.    XRD pattern of $RE_2CuO_4$ films prepared by MOD. All peaks can be indexed to the (00$l$) reflections of the $Nd_2CuO_4$ structure, indicating that the films are single-phase $T'$ and also single-crystalline.

Figure 2.    Temperature dependences of resistivity for $Sm_2CuO_4$ films (A and B). Film A was fired at 900°C for 1 h in $P_{O2}$ = 1 atm followed by reduction at 750°C for 10 min whereas Film B was fired at 850°C for 1 h in $P_{O2}$ = 2.8 x $10^{-3}$ atm followed by reduction at 440°C for 10 min.

Figure 3.    Reduction dependence of $\rho$-$T$ for $Pr_2CuO_4$ films: (a) varying the reduction temperature ($T_{red}$) with the reduction duration ($t_{red}$) fixed at 10 min, (b) varying $t_{red}$ with $T_{red}$ fixed at 450°C. The films were fired with identical conditions (850°C for 1h in $P_{O2}$ = 2.8 x $10^{-4}$ atm).

Figure 4.    Plots of $T_c$ and $\rho$(300 K) versus $c_0$ for all $Pr_2CuO_4$ films prepared by MOD. Upper ($T_c$-vs-$c_0$): open and filled circles stand for $T_c^{onset}$ and $T_c^{end}$. Lower ($\rho$(300 K)-vs-$c_0$): crosses stand for films not showing superconductivity, open and filled circles stand for films showing superconductivity with and without zero resistance, respectively. Data A - F are from the corresponding films in Fig. 5.

Figure 5.    Reduction dependence of $R_H$ for $Pr_2CuO_4$ films. The films are labeled from A to F according to the reduction strength, which are indicated in Fig. 4. $R_H$ goes up with the change of sign from negative to positive with reduction.

Table I. Comparison of $T_c$ between undoped $RE_2CuO_4$ and Ce-doped $RE_{1.85}Ce_{0.15}CuO_4$.[*]

| $RE$ | $T_c$ of $RE_2CuO_4$ [K] | $T_c$ of $RE_{1.85}Ce_{0.15}CuO_4$ [K] |
|---|---|---|
| La | – | 18.5[**] |
| Pr | 31.5 | 25.0 |
| Nd | 32.5 | 24.0 |
| Sm | 27.5 | 19.0 |
| Eu | 27.5 | 12.0 |
| Gd | 19.0 | 0 |

[*] $T_c$'s of $RE_{1.85}Ce_{0.15}CuO_4$ are taken from ref. [22].
[**]The highest $T_c$ of $La_{2-x}Ce_xCuO_4$ is 30.5 K at $x = 0.08$ [23].

Fig. 1 Matsumoto et al.

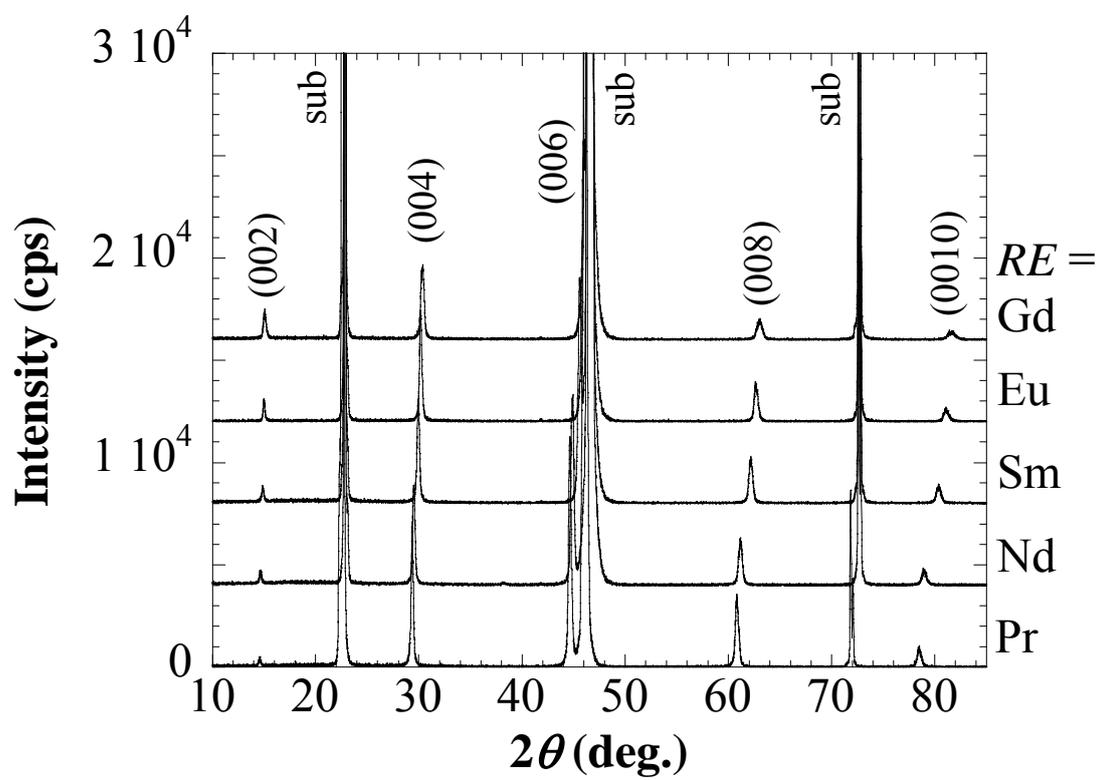

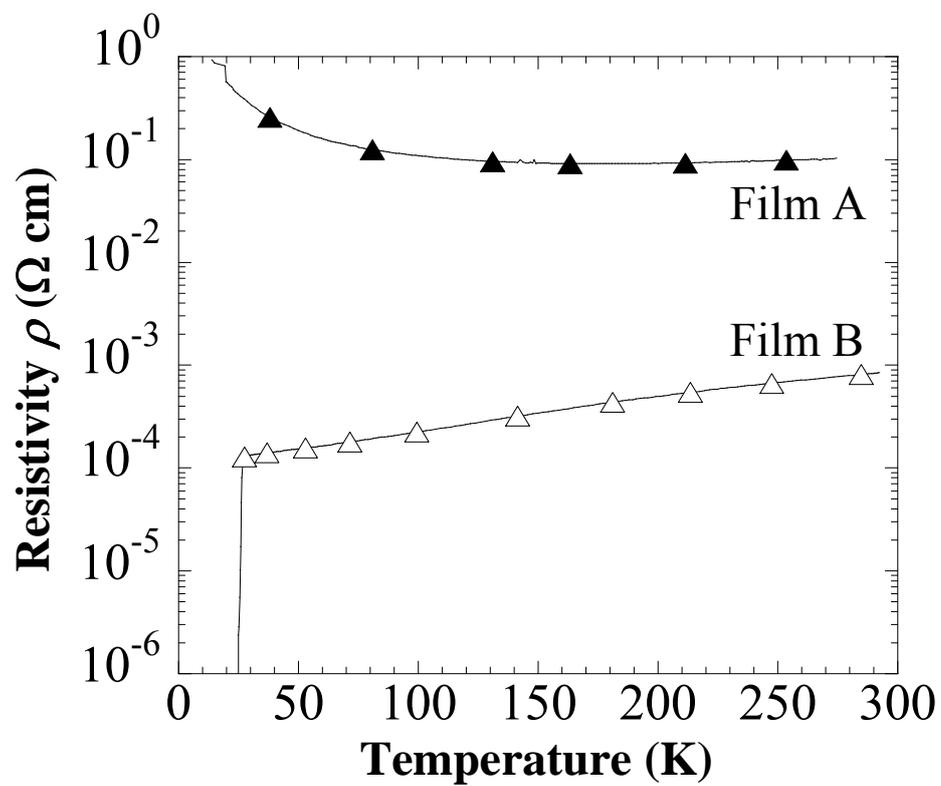

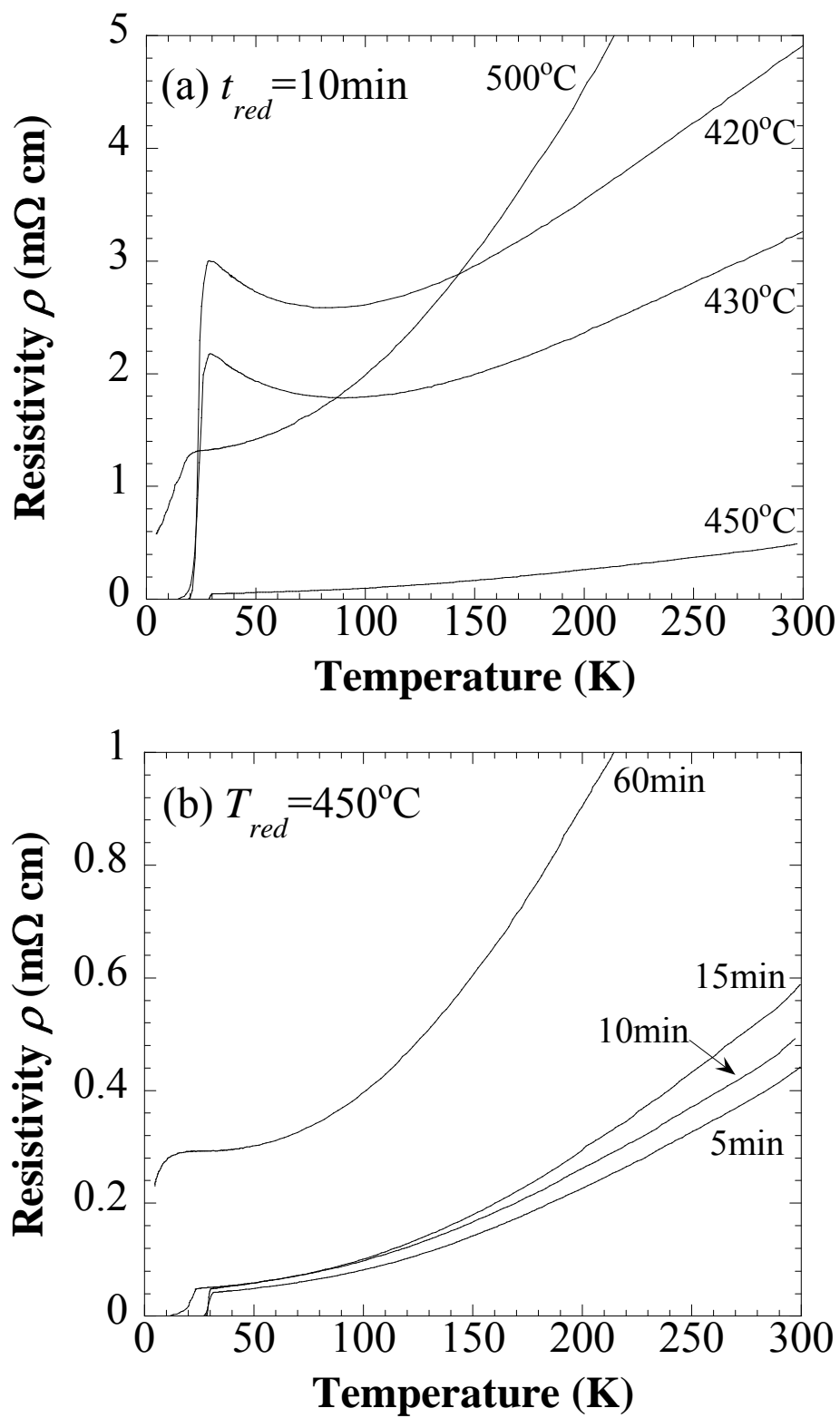

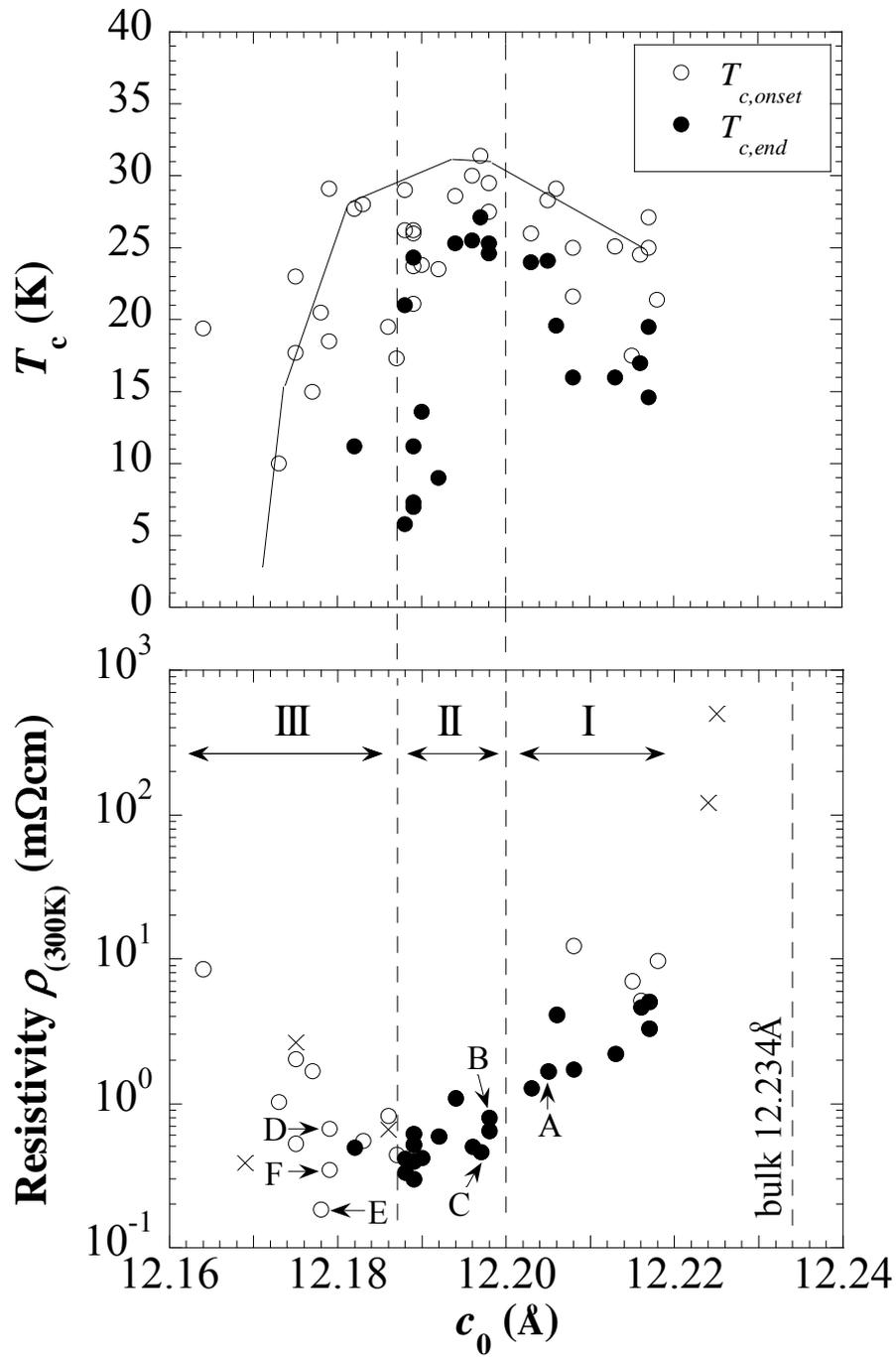

Fig 5 Matsumoto et al.

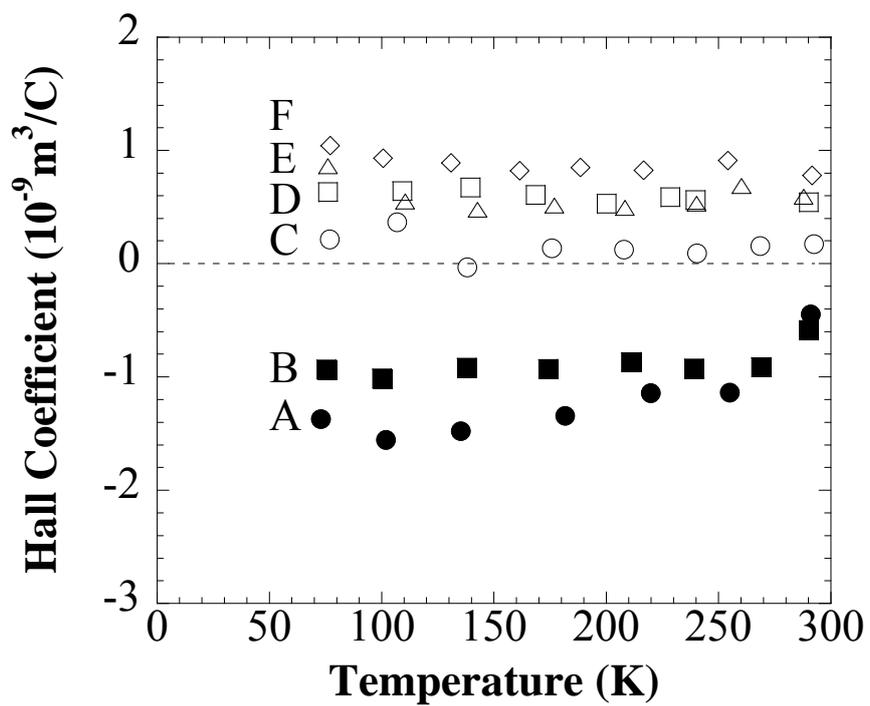